\documentclass[aps,prb,preprint,groupedaddress]{revtex4-1}
\usepackage{graphicx}
\begin{document}
\title{Direct Observation of Ferromagnetic State in Gold Nanorods Probed using Electron Spin Resonance Spectroscopy}
\author{Yuji Inagaki}
\email[inagaki.yuji.318@m.kyushu-u.ac.jp]{}
\author{Tatsuya Kawae}
\affiliation{Department of Applied Quantum Physics, Faculty of Engineering, Kyushu University, Fukuoka 819-0395, Japan}
\author{Natsuko Sakai}
\author{Yuta Makihara}
\affiliation{Department of Material Physics and  Chemistry, Faculty of Engineering, Kyushu University, Fukuoka 819-0395, Japan}
\author{Hiroaki Yonemura}
\author{Sunao Yamada}
\affiliation{Department of Applied Chemistry, Faculty of Engineering, Kyushu University, Fukuoka 819-0395, Japan}

\begin{abstract}
X-band electron spin resonance (ESR) spectroscopy has been performed for gold nanorods (AuNRs) of four different sizes covered with a diamagnetic stabilizing component, cetyltrimethylammmonium bromide. The ESR spectra show ferromagnetic features such as hysteresis and resonance field shift, depending on the size of the AuNRs. In addition, the ferromagnetic transition is indicated by an abrupt change in the spectra of the two smallest AuNRs studied. A large $g$-value in the paramagnetic region suggests that the ferromagnetism in the AuNRs originates from strong spin-orbit interaction.\end{abstract}
\maketitle
\section{Introduction}
In the past decade, rod-shaped gold nanoparticles, called Au nanorods (AuNRs), have attracted much attention because of their potential applications in sensing, imaging, and in vivo photothermal cancer therapy\cite{appl1, appl2, appl3, cancer}. These applications are based on the inherent tunable optical properties of the AuNRs by changing the aspect ratio  $a_{r}$ defined by the ratio of the length to the diameter \cite{opt}. In contrast, there have been few reports thus far concerning the magnetic properties of AuNRs\cite{mag-aunr}, although there have been many studies on the magnetism of spherical Au nanoparticles (AuNPs)\cite{thiol, hori, au-ag-cu, neutron, soft-ferro, jacs}.

Recently, Yonemura $et$ $al.$ examined the effects of magnetic processing and observed the magnetic orientation and the side-by-side aggregation of AuNRs/poly(styrenesulfonate) composites, AuNRs with three different aspect ratios ($a_{r}$ = 8.3, 5.1, and 2.5), and Au nanowires under a magnetic field\cite{yonemura, yonemura2, yonemura3}, which implies that the magnetic response of AuNRs is positive, although bulk Au has been believed to be diamagnetic for a long time. However, the detailed properties and origin of magnetism in AuNRs is not yet well understood.

Generally, magnetic properties have been investigated through magnetization measurements using highly sensitive magnetometers, such as a commercial SQUID magnetometer. In nano-sized particles coated with a stabilizing polymer, however, it is difficult to extract the net magnetization precisely owing to the superposition of a large amount of diamagnetic components\cite{error}. We focus on the electron spin resonance (ESR) technique to eliminate the signal from the stabilizing polymer because ESR is insensitive to diamagnetism. Moreover, the skin depth of Au at the X-band microwave region ($\sim$ 9 GHz) is about 0.8 $\mu$m, which is much larger than the size of AuNRs in the present study. These features indicate that ESR is a powerful tool for investigating the magnetic properties of AuNRs. 

In the present study, X-band ESR measurements were performed in detail for AuNRs of four different sizes, where $a_{r}$ was varied from 2.5 to 8.3. Ferromagnetic features were observed in all AuNRs. In addition, ferromagnetic transitions were clearly detected in the two smallest AuNRs studied. From the ESR parameters deduced from the conduction electron spin resonance (CESR) at high temperatures, strong spin-orbit interaction and resultant large effective mass of conduction electrons are pointed out for the origin of the magnetism in the AuNRs on the basis of Elliott-Yafet theory.

\section{Experimental}
AuNRs studied here are named as s-, s$^{\prime}$-, m-, and l-AuNR from small to large values of $a_{r}$. All AuNRs are prepared by the soft template method using cetyltrimethylammmonium bromide (CTAB)\cite{gosei}. The s- and s$^{\prime}$-AuNR are prepared using a reducing agent, triethylamine, with or without acetone. The m-AuNR is prepared using a combination of chemical reduction and photoreduction. The l-AuNR is prepared using two kinds of reducing agents, sodium borohydride and triethylamine.
Typical TEM images of AuNRs are given in ref.\cite{yonemura}. X-ray fluorescence and CHN elemental analyses confirm that AuNRs contain no contamination of other magnetic metals, as listed in Table~\ref{aunr}, which ensures that the magnetic response in the present study originates from the Au atoms in AuNRs. ESR measurements are performed using an X-band microwave system (JEOL ES-SCEX) equipped with a continuous-He-flow-type cryostat (Oxford ESR910) operating down to $T$ $\sim$ 5 K.
\begin{table*}
\caption{\label{aunr}%
Dimensions, aspect ratio, estimated content of Au, CTAB and other metals for each AuNR
}
\begin{ruledtabular}
\begin{tabular}{ccccccc}
\textrm{}&
\textrm{size (nm)}&
\textrm{$a_{r}$}&
\textrm{Au(\%)}&
\textrm{CTAB(\%)}&
\textrm{Ag(\%)}\\
\colrule
s-AuNR & 4.6$\phi \times$11.6 & 2.5 & 3.6 & Br:22.6, C:59.3, H:10.8, N:3.7 & -\\
s$^{\prime}$-AuNR & 5.0$\phi \times$20.0 & 4.0 & 41.3 & Br:16.7, C:31.4, H:5.8, N:1.9 & 2.9\\
m-AuNR & 7.2$\phi \times36.6$ & 5.1 & 3.1 & Br:22.9, C:59.4, H;11.0, N:3.6 & -\\
l-AuNR & 7.7$\phi \times$63.8 & 8.3 & 1.9 & Br:22.7, C:60.6, H:11.2, N:3.7 & -\\
\end{tabular}
\end{ruledtabular}
\end{table*}

\section{Results and Discussion}
\begin{figure}
\includegraphics[width=80mm]{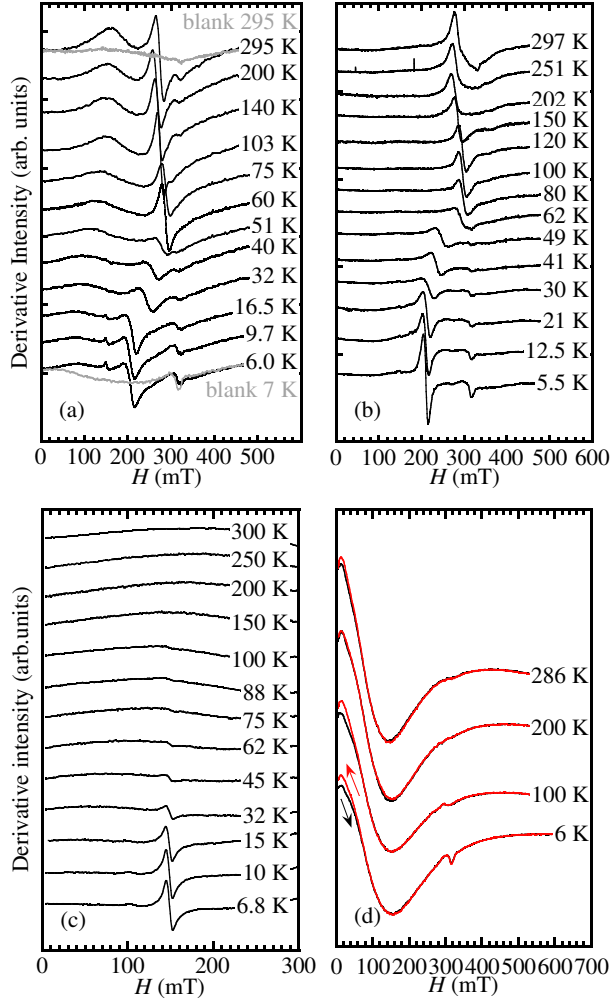}
\caption{\label{f1} (Color online) Temperature dependence of ESR spectra in (a) s-AuNR, (b) s$^{\prime}$-AuNR (c) m-AuNR and (d) l-AuNR. The absorption  around 320 mT is due to instrumental background, as observed from the blank spectra recorded at 295 K and 7 K shown by the gray lines in (a).}
\end{figure}

Figures~\ref{f1}(a) and~\ref{f1}(b) show the temperature dependence of ESR spectra observed for s-AuNR and s$^{\prime}$-AuNR, respectively.
 Two spectra plotted with gray lines in Fig.~\ref{f1}(a) correspond to the blank signals recorded at 7 K and 295 K, which originate from the background from instruments including the sample holder (quartz tube) and the CTAB. This indicates that the absorptions at around 320 mT are not intrinsic signals from the AuNR. At a glance, it seems that the spectra from the two AuNRs show quite similar temperature dependences. However, on closely observing both spectra, several differences can be discerned. At room temperature, the ESR spectrum of s-AuNR consists of two components: a sharp absorption centered at s$_{1}$ = 275 mT and a broad one around s$_{2}$ = 230 mT, which is strongly suppressed when the temperature is decreased. In contrast, the spectra of s$^{\prime}$-AuNR have no broad component and show only a sharp absorption centered at s$^{\prime}_{1}$ = 289 mT. Thus, the sharp absorptions at s$_{1}$ and s$^{\prime}_{1}$ are regarded as intrinsic properties of AuNRs, and their temperature variations are examined below.

\begin{figure}
\includegraphics[width=60mm]{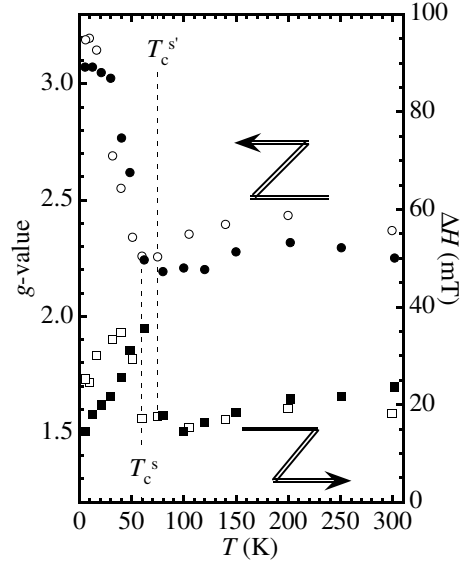}
\caption{\label{f2} Temperature dependence of $g$-value (circle) and line width $\Delta$$H$(square). Open and solid symbols represent the results for s-AuNR and s$^{\prime}$-AuNR, respectively.}
\end{figure}

The temperature dependences of $g$-values estimated from resonance fields and line widths  $\Delta$$H$ for s$_{1}$ and s$^{\prime}_{1}$ are summarized in Fig. ~\ref{f2}.
The resonance field and line width are nearly independent of temperature for both samples above $\sim$80 K, while the spectra show a drastic shift of the resonance field and broadening of the width below 55 K for s-AuNR and 75 K for s$^{\prime}$-AuNR. The $g$-value of s-AuNR (s$^{\prime}$-AuNR) increases with decreasing temperature and reaches 3.19 (3.07) at the lowest temperature $T$ $\sim$ 6 K. For s-AuNR (s$^{\prime}$-AuNR), the line width shows a maximum at approximately 40 K (60 K), which is followed by a gradual narrowing with decreasing temperature. This series of behaviors is a typical feature observed in the magnetic ordering process. Hence, we conclude that the resonance field shift and broadening of the width in s- and s$^{\prime}$-AuNRs are caused by magnetic ordering with transition temperatures of $T_{c}^{s}$ $\sim$ 55 K and $T_{c}^{s^{\prime}}$ $\sim$ 75 K, respectively. 

The results for m- and l-AuNRs are depicted in Figs.~\ref{f1}(c) and~\ref{f1}(d), respectively. In m-AuNR, the ESR spectrum cannot be observed in the high-temperature region. When the temperature is decreased below $T$ $\sim$ 100 K, a narrow absorption with $\Delta$$H$ = 8 mT appears at around 150 mT. The intensity of the spectrum grows with decreasing temperature, while both the resonance center and line width are almost independent of temperature down to $T$ = 6.8 K. This temperature dependence will be discussed later.

In contrast, l-AuNR shows broad ESR spectra in the entire temperature range, as shown in Fig.~\ref{f1}(d). The resonance field at $\sim$70mT in l-AuNR is the lowest among all AuNRs, while the line width of  $\sim$130 mT is the broadest. It is significant that a hysteresis emerges in all the spectra of l-AuNR below  $\sim$60 mT in the magnetic field sweep, as indicated by arrows. A hysteresis between magnetizing and demagnetizing processes is a characteristic feature of a ferromagnet with domain structures and/or magnetic anisotropy\cite{hysteresis1, hysteresis2}. Note that the domain structure and anisotropy give rise to the line broadening in ESR spectra. In other words, these features observed in l-AuNR are well explained by assuming a large scale ferromagnet.

\begin{figure}
\includegraphics[width=80mm]{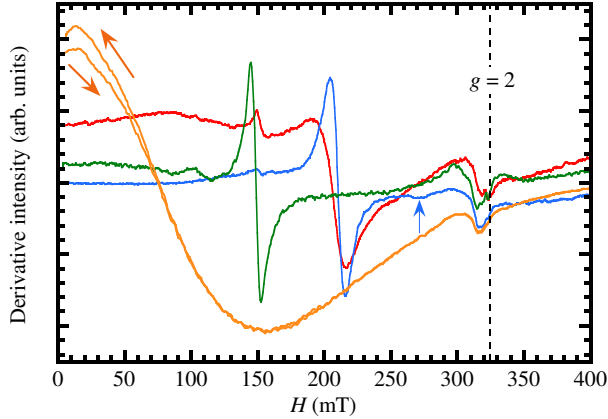}
\caption{\label{f3}  (Color online) ESR spectra of all AuNRs at T $\sim$ 6K.  Red, blue, green and orange colors correspond to s-, s$^{\prime}$-, m- and l-AuNR, respectively.  The vertical dotted line indicates the resonance field for $g$=2.}
\end{figure}

Next, we discuss the systematic resonance shift in all AuNRs. The spectra at the lowest temperature $T$ $\sim$ 6 K are plotted together in Fig.~\ref{f3} to make a comparison between all AuNRs.
 In a ferromagnet with a cylindrical shape like the present AuNRs, ferromagnetic resonance (FMR) occurs at a resonance field that depends on the demagnetizing form factor owing to the aspect ratio $a_{r}$, as well as on the magnitude of moment  $M_{s}$ and anisotropy $K$. In fact, the magnitude of shift increases with $_{a}$r in the present systems; the shift is estimated to be approximately 248, 176, 114, and 118 mT for l-, m-, s$^{\prime}$-, and s-AuNR, respectively, which corresponds to the order of $a_{r}$ except for s- and s$^{\prime}$-AuNR. The reversal between them may be caused by the resonance shift dominated by an effective anisotropy field represented by $K/M_{s}$. Thus, a small anisotropy $K$ can give rise to large effective field if $M_{s}$ is small. As for the anisotropy, the following qualitative discussion can be made. In randomly oriented ferromagnetic species, a powder pattern is expected in the FMR spectrum. Although a clear powder pattern was not recorded for all AuNRs, finite ESR intensity can be seen like a tail in a higher field range than the main absorption peak, as represented in l-AuNR. It is difficult to refer about other AuNRs because of the superposition of background signal at around 320 mT. However, a small intensity like shoulder is visible at about 270 mT in s$^{\prime}$-AuNR, as indicated by the vertical arrow in Fig.~\ref{f3}. Such FMR spectra correspond to the case with a negative anisotropy in the cubic symmetry\cite{fmr}. Detailed frequency dependence of ESR is required to obtain further information about the FMR parameters $M_{s}$ and $K$.

The observed systematic shift, which is generally not expected in other antiferromagnets, paramagnets, and ferromagnets with a spherical shape, provides further evidence to the existence of ferromagnetic states at low temperatures in all the present AuNRs. On the basis of the hysteresis and systematic shift of the resonance field, it is reasonable to consider that all AuNRs are in the ferromagnetic state at low temperatures. To our knowledge, this is the first time that a ferromagnetic state has been found in AuNRs.

In the final part of this Letter, we examine the ESR spectra in the paramagnetic region. The spectra of s- and s$^{\prime}$-AuNRs above $T_{c}$ can be understood as CESR ones. In the CESR region, $g$-values are estimated to be 2.34$\pm$0.09 and 2.26$\pm$0.06 for s- and s$^{\prime}$-AuNRs, respectively. These values are considerably larger than the value of 2.11 reported for bulk Au but comparable to that of 2.26$\pm$0.02 for small particles of Au with a mean diameter of 3 nm\cite{bulk-esr, nano-esr}. For CESR, the difference  $\Delta$$g$ between the $g$-value of real metals and that of ideal free electrons (2.0023) gives an approximate value of the spin-orbit interaction through the equation $\Delta$$g$$\sim$ $\lambda$/$\Delta$$E$. Here, $\lambda$ is the spin-orbit coupling constant and $\Delta$$E$ is the difference in energy between the 6s band and the nearest 5d band\cite{yafet}. Thus, the large $\Delta$$g$ observed in the present measurements indicates the large contribution of the orbital moment in the 5d band to conduction electrons in the 6s band, which governs the magnetic properties of AuNRs and leads to the ferromagnetic state at low temperatures. A strong spin-orbit coupling was also confirmed in recent studies of X-ray magnetic circular dichroism in bulk Au as well as AuNPs\cite{xmcd-bulk, xmcd-nano}. 

The strong spin-orbit interaction also causes a significant broadening of line width. The line width of CESR in metals is closely related to the spin-lattice relaxation time, i.e., the spin flip rate by phonons. Accordingly, $\Delta$$H$ varies linearly with temperature. Therefore, at high temperatures, CESR is hardly detected in not only bulk metals but also m-AuNR.  This is a reason why we could not observe the ESR in m-AuNR above 100 K.  In contrast, narrow ESR absorptions are obtained for both s- and s$^{\prime}$-AuNRs, which show rather gentle temperature dependences without a marked increase in the width. These features may be explained by considering the system size of AuNRs. In small systems, there exists a lower limit of phonon mode frequency given by $\nu$ = $v_{s}$/2$L$, where $L$ is the largest dimension of the system and $v_{s}$ is the speed of sound in the crystal; this results in an increase in the spin-lattice relaxation time of conduction electrons\cite{honig}. As a result, narrow absorptions are observed in s- and s$^{\prime}$-AuNRs. This scenario is qualitative and does not account for the nearly equal line widths of s-AuNR and s$^{\prime}$-AuNR; other factors should be considered for the quantitative explanation.

\begin{figure}
\includegraphics[width=60mm]{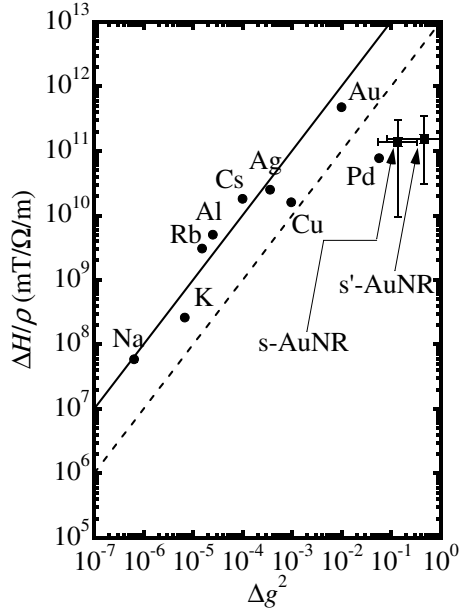}
\caption{\label{f4} Beuneu-Monod plot, which connects $\Delta$$g^{2}$ and $\Delta$$H$/$\rho$, for pure metals given in ref.~\cite{beuneu} and for two systems in the present study. The value $\Delta$$H$/$\rho$ of most metals differs from the fit, which is indicated by the solid and dashed lines, by less than one order of magnitude.}
\end{figure}

Nevertheless, owing to the narrow line widths, we could successfully observe CESR of AuNR, which allows identification of the characteristic feature of the present CESR results through sorting with other metals in the so-called Beuneu–Monod plot shown in Fig.~\ref{f4}\cite{beuneu, monod}.
 The Beuneu–Monod plot is an empirical plot that connects CESR parameters $\Delta$$g$ and $\Delta$$H$ for pure metals via $\Delta$$H$/$\rho$ = $\alpha\Delta$$g^{2}$, where $\rho$ is the resistivity and $\alpha$ is a metal-dependent constant. Most metals follow a straight-line fit in the log-log plot of $\Delta$$H$/$\rho$ vs.  $\Delta$$g^{2}$ by less than one order of magnitude, as indicated by the solid and dashed lines in Fig.~\ref{f4}. The values used in this plot are taken at a temperature of approximately $T_{D}$/7, where $T_{D}$ is the Debye temperature. In the case of bulk Au, $T_{D}$/7 corresponds to 20 K. The CESR parameters for the two systems in the present study (s- and s$^{\prime}$-AuNR) are not available for such a low temperature, which is less than $T_{c}$. Thus, we plot the estimate taken in the temperature range between $T_{c}$ and room temperature, while the resistivity is fixed at the bulk value at 20 K. Error bars for both axes originate from the temperature variation of $\Delta$$g$ and $\Delta$$H$. In the plot, the two data points corresponding to s- and s$^{\prime}$-AuNR deviate significantly from those of other metals, but they are located near the point corresponding to Pd. It is well known that Pd is close to satisfying the Stoner criterion even in a bulk form. Indeed, ferromagnetism is realized by the downsizing of Pd\cite{palladium1, palladium2, ienaga-pd}. Accordingly, it is confirmed that the two AuNRs are also close to the ferromagnetic state. 

To reproduce the results for s-AuNR, s$^{\prime}$-AuNR, and Pd, a smaller $\alpha$ is needed. This suggests that the low value of $\alpha$ is closely related to the realization of ferromagnetism. According to the Elliott-Yafet theory, the coefficient $\alpha$ is given by $ne^{2}$/$\gamma$$m^{\ast}$, where $n$, $e$, and $m^{\ast}$ are the density, charge, and effective mass of the conduction electron, respectively, and $\gamma$ is the magnetomechanical ratio. Therefore, the origin of the deviation caused by the narrow $\Delta$$H$ and the large $\Delta$$g$ is suggested to be an enhancement of the effective mass. This is a reasonable conclusion because a large effective mass is realized in a typical ferromagnet of Ni, as revealed recently by high-resolution angle-resolved photoemission studies\cite{ni}. 

In summary, we performed ESR measurements for AuNRs of four sizes with different aspect ratios. The ESR spectra at low temperatures are explained in the context of ferromagnetic resonance for all AuNRs. Detailed frequency dependence of FMR measurements will enable further quantitative discussion on the size- and shape-dependent magnetism of AuNRs. In the two smallest AuNRs, we detected ferromagnetic transitions in the ESR spectra at $T$ $\sim$ 60 K, which offers an opportunity to explore the critical behavior of the phase transition in nano-rod systems. The CESR above  $T_{c}$ suggests that the strong spin-orbit interaction is responsible for ferromagnetism in the AuNR systems.

\begin{acknowledgments}
The authors are grateful to T. Asano and T. Sakurai for the help of ESR experiments and useful discussion. This work was partially supported by a Grant-in-Aid for Scientific Research, No.23540392, No.25220605 and No.25287076. The authors thank to Dr. Daigou Mizoguchi (Dai Nippon Toryo Co. Ltd.) for providing them with four different types of AuNRs. The authors are also grateful to the Center of Advanced Instrumental Analysis, Kyushu University for X-ray fluorescence, CHN elemental analyses and the use of a TEM apparatus.
\end{acknowledgments}
\bibliography{AuNR-Inagaki-Nov2014}
\end{document}